\documentclass{ws-ijmpa}

\begin{document}

\markboth{Akira Ukawa}
{Hadron spectrum from Lattice QCD}

\catchline{}{}{}{}{}
%
%%%%%%%%%%%%%%%%%%%%%%%%%%%%%%%%%%%%%%%%%%%%%%%%%%%%%%%%%%%%%%%%%%%%

\title{Hadron Spectrum from Lattice QCD
}

\author{\footnotesize AKIRA UKAWA}

\address{Center for Computational Sciences, University of Tsukuba\\
Tsukuba, Ibaraki 305-8577, Japan
}

\maketitle

\pub{Received (Day Month Year)}{Revised (Day Month Year)}

\begin{abstract}
A brief review is given of the lattice QCD calculation of the hadron spectrum. 
The status of current attempts toward inclusion of 
dynamical up, down and strange 
quarks is summarized focusing on our own work.  
Recent work on the possible existence of pentaquark states are 
assessed.  We touch upon the PACS-CS Project for building our next machine for 
lattice QCD, and conclude with a near-term physics and machine prospects.


\end{abstract}

\section{Introduction}	%) A SECTION HEADING

The spectrum of hadrons is a fundamental entity for lattice QCD 
elucidation of the dynamics of the strong interactions.  Calculating 
the spectrum of standard hadrons such as pion, nucleon, $J/\psi$, $\Upsilon$, 
and $D$'s and $B$'s provides a basic mean for confirmation of the theory and a 
benchmark for the lattice calculational methods.  Finding the spectrum of 
non-standard hadrons, {\it e.g.,} glueballs, hybrids and multi-quark states, 
offers chances for predictions and thereby further tests of QCD.  For these 
reason the hadron spectrum has been repeatedly and routinely calculated 
over the years with increasing accuracy and sophistication. 

At present, the focus of spectrum calculations and many other 
subjects with phenomenological impact centers around dynamical 
simulations including the sea quark effects of all three light quarks, 
up, down and strange.  There are two major attempts, one using the 
staggered quark action\cite{milc} and the other using the Wilson-clover 
quark action\cite{cppacs-jlqcd}.  In addition, a third attempt with the 
domain-wall quark action, made possible with the commissioning of QCDOC
with 10Tflops-class capability, are beginning\cite{uk-rbrc}. 
In Sec.~\ref{sec:spectrum} we discuss results from our own attempt on the 
light hadron spectrum and quark masses, and recent preliminary results 
on heavy quark systems. 

Search for possible 
pentaquark states has occupied a lot of experimental and theoretical effort. 
A number of lattice QCD simulations have also been made to see if there is 
signal indicating their theoretical presence.  A crucial and special issue
with the pentaquark concerns distinguishing bound or resonance states from 
scattering states.  Recent lattice studies focused on applying theoretical 
criteria on this point, and we shall review them in Sec.~\ref{sec:pentaquark}. 
We conclude with a brief summary in Sec.~\ref{sec:conclusion}.

\section{$N_f=2+1$ lattice QCD simulations with the Wilson-clover quark action}
\label{sec:spectrum}

There are two major lattice QCD research collaborations in the Tsukuba 
area in Japan, the CP-PACS Collaboration based at University of 
Tsukuba and the JLQCD Collaboration based at KEK.  Since 2001 the two 
collaborations have jointly pursued a project to carry out 
simulations including dynamical up, down and strange quarks.  The strategy 
adopted, and the necessary preparations carried out prior to actual runs, 
is as follows: (i) use Iwasaki RG-improved gluon action to span a 
range of lattice spacing toward coarse lattices and avoid the artificial 
phase transition observed for the plaquette gluon action\cite{artifact}, 
(ii) use Wilson-clover quark action with a fully $O(a)$-improved clover 
coefficient calculated with the Shcr\"odinger functional methods for 
three dynamical flavors\cite{csw}, and (iii)apply the polynomial HMC 
to handle the strange quark\cite{phmc} in addition to the standard HMC 
for the up and down quarks which are treated as degenerate.      

In the table below we list the parameters of our simulations.  
Runs are made at three values of lattice spacing equally spaced in $a^2$.  
At each lattice spacing, five values are taken for the degenerate up and 
down quark mass in the range $m_{PS}/m_V\approx 0.6-0.8$, and two values 
for the strange quark mass at $m_{PS}/m_V\approx 0.7$. 
Hadron masses calculated for these quark masses are fitted with a 
general quadratic polynomial of VWI quark masses 
$m^{VWI}=(K^{-1}-K^{-1}_c)/2$, and are extrapolated to the physical point 
defined either by the experimental $\pi, \rho$ and $K$ meson masses 
($K$-input) or $\pi, \rho$ and $\phi$ meson masses ($\phi$-input).  
The agreement of the lattice spacing determined by the two types of input 
as seen in Table \ref{tab:runs} provides a simple but 
important check on the internal consistency of the $N_f=2+1$ calculation, 
which is not realized in quenched and $N_f=2$ simulations. 

\begin{table}[b]
\begin{center}
\caption{Simulation parameters.  The production run at $\beta=2.05$ is still in progress.}
\caption{test}
\label{tab:runs}
\begin{tabular}{ccccc}
\multicolumn{5}{l}{Parameters of our simulations}\\
\hline
$\beta$ & size & $a$ [fm] ($K$-input) & $a$ [fm] ($\phi$-input) & trajectory \\
\hline
$1.83$ & $16^3\times32$ & $0.1222(17)$ & $0.1233(30)$ & $7000 - 8600$ \\
$1.90$ & $20^3\times40$ & $0.0993(19)$ & $0.0995(19)$ & $5000 - 9200$ \\
$2.05$ & $28^3\times56$ & $0.0758(48)$ & $0.0755(48)$ & $3000 - 4000$ \\
\hline
\end{tabular}
\end{center}
\end{table}

\begin{figure}
\begin{center}
\includegraphics[width=120mm,angle=0]{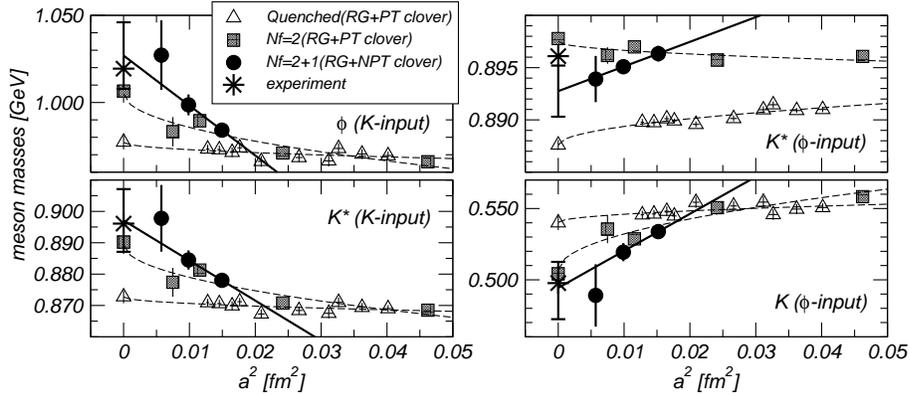}

\caption{Continuum extrapolation of meson masses,
         compared with those for quenched and $N_f=2$ 
         QCD~\protect\cite{Ali_Khan:2000-1}.
         Note that the quenched and $N_f=2$ simulations are made with 
         the one-loop perturbatively $O(a)$-improved clover action.
         Thus extrapolations are made linearly in $a$.}
\label{fig:hyperfine}
\end{center}
\end{figure}

\subsection{Light quark sector} 

In Fig.~\ref{fig:hyperfine} we plot the continuum extrapolation of meson 
masses in quenched(triangles), $N_f=2$(squares) and the present $N_f=2+1$(circles) simulations\cite{ishikawa}.   
The solid lines for the $N_f=2+1$ data are pure quadratic fits to the 
results for the two coarse lattice spacings for which runs and measurements 
have been completed. The agreement with experiment in the continuum limit is 
encouraging, but we need further data at the finest lattice spacing for 
a solid conclusion.          

\begin{figure}
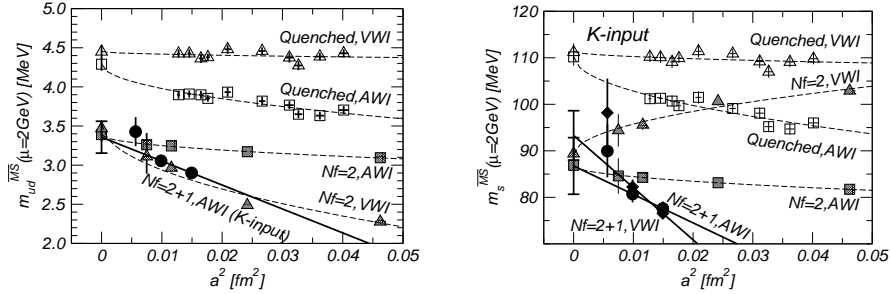

\begin{center}
\includegraphics[width=55mm,angle=0]{./Figure/Mud_vs_a2_k-input.eps}
\hspace{0.5cm}
\includegraphics[width=55mm,angle=0]{./Figure/Ms_vs_a2_k-input.eps}
\caption{Continuum extrapolations of the up, down and strange quark masses
         obtained with the $K$-input.
         The data at the finest lattice is not included in
         the continuum extrapolations.
         For comparison, results for quenched and $N_f=2$ QCD are
         overlaid.}
\label{fig:uds_quark_masses}
\end{center}
\end{figure}

Figure \ref{fig:uds_quark_masses} shows the continuum extrapolation for the 
light quark masses\cite{ishikawa}.  
Values sizably smaller than the quenched estimate as 
has been strongly suggested in the previous $N_f=2$ simulations are confirmed. 
Again, we wait completion of the analyses at the finest lattice spacing to 
quote the final result for light quark masses in the continuum limit. 

An important issue with the analyses in the light quark sector concerns 
chiral extrapolation.  Since Wilson-clover quark action involves explicit 
chiral symmetry breaking, the chiral behavior of physical quantities 
deviate from that in the continuum.  The Wilson chiral perturbation 
theory\cite{WChPT} provides the procedure to work out the chiral behavior 
for finite lattice spacings for Wilson-type actions, 
and work is under way to calculate the 
consequence of the approach for the spectroscopic quantities including 
pseudo scalar meson masses, decay constants, and quark masses on the one 
hand, and to analyze data based on those results.

\subsection{Heavy quark sector} 

Calculating the physical quantities of hadrons containing heavy quarks 
is important to constrain the parameters of the Standard Model 
and to help explore physics at finer scales.
A serious obstacle is large systematic errors of form $O(m_Ha)$ where $m_H$ 
denotes the heavy quark mass, which are large at currently accessible lattice 
spacings.  A standard lore for overcoming this problem is to resort to 
heavy quark effective theory approach such as NRQCD.  We wish to 
pursue a different approach in which parameters of the quark action are tuned 
demanding Lorentz invariance of on-shell quantities for arbitrary large 
$m_Ha$ to $O(a)$\cite{rhq}.  This approach allows the continuum 
extrapolation to be taken in contrast to effective theory approaches. 

\begin{figure}[t]
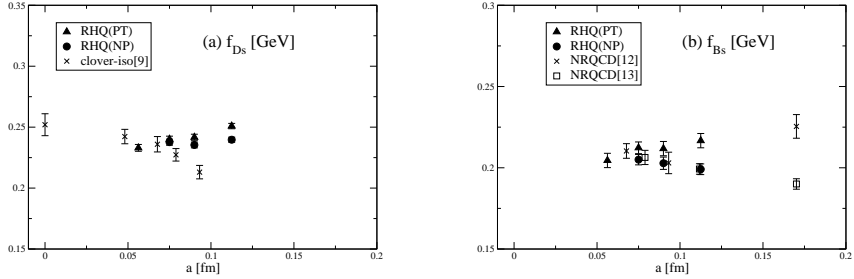

\begin{center}
\includegraphics[width=50mm,angle=0]{Figure/fcs.eps}
\hspace{1cm}
\includegraphics[width=50mm,angle=0]{Figure/fbs.eps}
\end{center}
\caption{Decay constant for (a) $D_s$ and (b) $B_s$ in quenched QCD from 
relativistic heavy quark approach\protect\cite{kuramashi} as compared to 
those of effective theories\protect\cite{fd_clover,fb_nrqcd,fb_jlqcd}.}
\label{fig:fps}
\end{figure}

We have been testing this approach for charm and bottom systems 
on $N_f=2$ and quenched configurations prior to applying it to  
$N_f=2+1$ QCD. In  Fig.~\ref{fig:fps} we show results for the decay 
constant of $D_s$ and $B_s$ mesons in quenched QCD\cite{kuramashi}
and compare them with those of 
effective theory approaches\cite{fd_clover,fb_nrqcd,fb_jlqcd}. 
Mild cutoff dependence of the present results and consistency 
at finite lattice spacings indicate success of our approach.  

\subsection{The next step}

At present the largest limitation in our data is a rather large value of the 
up and down quark mass $m_{ud}$.  In terms of the physical strange quark mass 
$m_s^{phys}$, it only goes down to $m_{ud}\approx m_s^{phys}/2$ while 
experimentally $m_{ud}\approx m_s^{phys}/25$.  We wish to reduce the 
value to at least $m_{ud}\approx m_s^{phys}/5$ or below in order to control 
the chiral behavior.  

This requires both an enhancement of the computing power and an improvement 
of the algorithm.  
The situations in both respects are very promising at present.  
For the latter, the domain-decomposition acceleration of 
HMC\cite{luscher} offers a very promising resolution. 
On the latter, in December 2004, Japanese Government formally approved 
our proposal to 
develop a massively parallel cluster at the Center for Computational 
Physics, University of Tsukuba.  We list in table below the 
specification of the cluster, which is named PACS-CS (Parallel Array 
Computer System for Computational Sciences).  The development is well 
under way, and the installation and start of operation are scheduled in 
June-July of 2006. 
We plan to combine these developments to break through the current 
limitations in our Wilson-clover program. 

\begin{table}[b]
\begin{center}
\begin{tabular}{ll|l}
\multicolumn{3}{l}{Design specification of PACS-CS}\\
\hline
Number of nodes & 2560&Node configuration\\
Peak performance & 14.3 Tflops &single LV Xeon 2.8GHz 5.6Gflops\\
Total memory & 5 TByte    &2 GByte memory 6.4GByte/s \\
Total disk space & 0.41 PByte &160 GByte$\times 2$(RAID1) local disk\\
Interconnect& $16\times 16\times 10$&Interconnect bandwidth\\
&3-dim hyper-crossbar &250MByte/s$\times 2$/link\\
OS  & Linux and SCore &750MByte/s$\times 2$/node\\
Programming & Fortran90, C, C++, MPI\\
\hline
System size & 59 racks & Estimated power 545 kW\\
\hline 
\end{tabular}
\end{center}
\label{tab:pacs-cs}
\caption{PACS-CS design specification}
\end{table}

\section{Lattice pentaquark search}
\label{sec:pentaquark}

A standard lattice QCD calculation of the mass of a hadron starts with 
a preparation of the operators having the desired quantum numbers and 
an examination of the large-time behavior of the two-point function to 
extract the ground state signal. 
With lattice pentaquark searches, we have to distinguish a possible 
pentaquark bound state or resonance from the nucleon-kaon scattering state. 
Hence, we need to disentangle at least two states in the correlation 
function, and we also have to distinguish a bound state/resonance from 
scattering states.  In addition, since the spin-parity of the pentaquark 
state is yet unknown, we have to explore over a large operator space. 

The initial lattice studies\cite{penta-first} did not explore these 
points in detail.  Subsequent calculations\cite{penta-since,fodor} addressed 
them using two techniques.  For multi-state analyses, 
a variational method using a set of operators and diagonalizing the normalized 
correlator matrix has been known for a long time.
This method has been employed in a number of recent pentaquark searches, and 
has been reasonably successful.  For distinguishing a scattering state, 
a basic strategy is to examine the spatial size dependence of the 
energy eigenvalues obtained in the multi-state analysis.  
A scattering state, if the interactions between the scattering hadrons are 
weak as in the case of the nucleon and kaon for the pentaquark case, would 
show a size dependence expected from the sum of two energies $\sqrt{m^2+(2\pi\ell/L)^2}$ with $\ell=1,2,3,\cdots$.  Another criteria is to look at the 
residue of the state in question in the two point function : 
$<O(t)O(0)>\stackrel{\to\infty}{\rightarrow}Z exp(-mt)$. 
For a scattering state we expect the overlap of the two particles
decreasing as $Z\propto 1/V$.  

\begin{figure}[t]
\begin{center}
\includegraphics[width=50mm,angle=0]{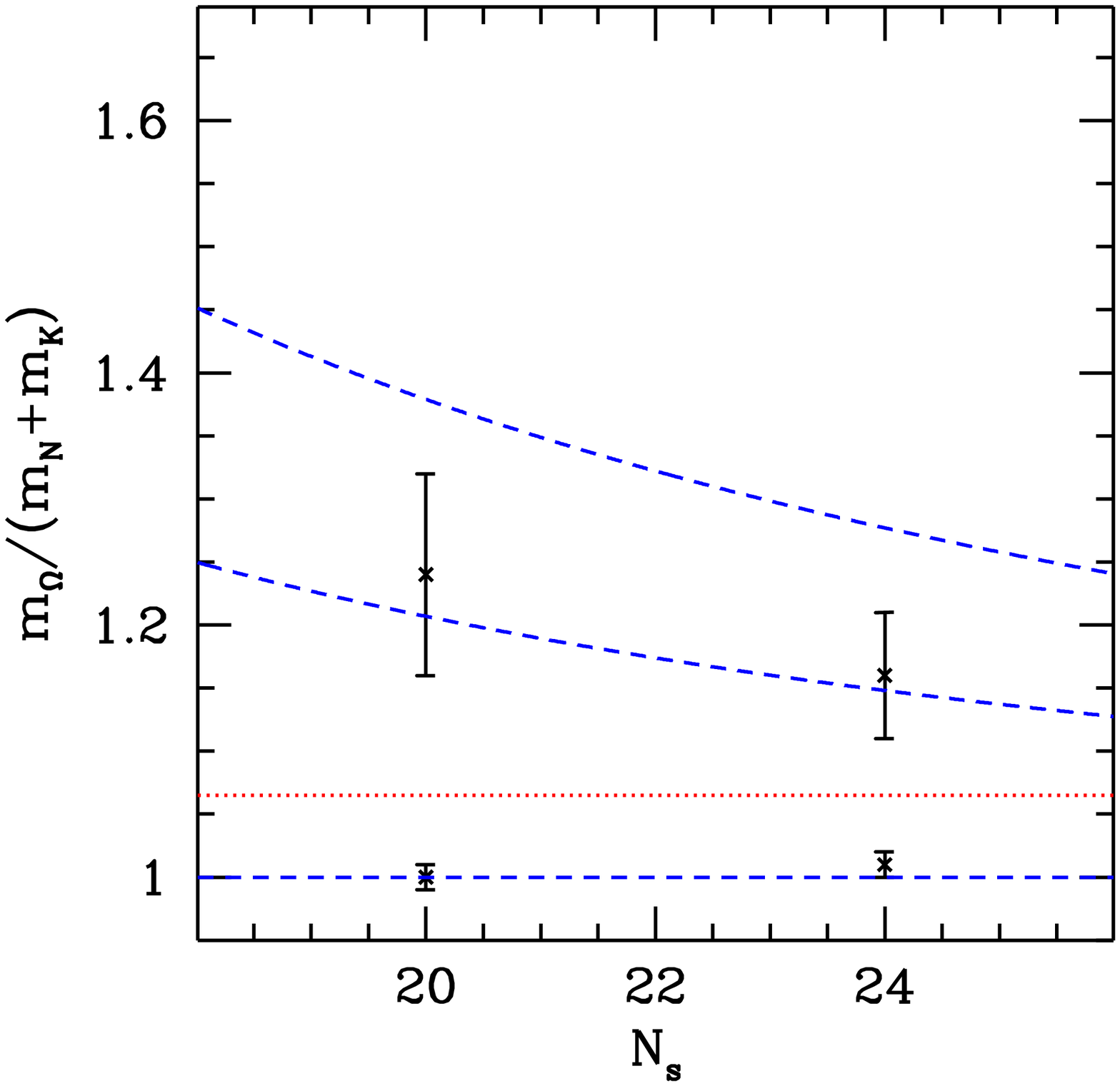}
\hspace{1cm}
\includegraphics[width=50mm,angle=0]{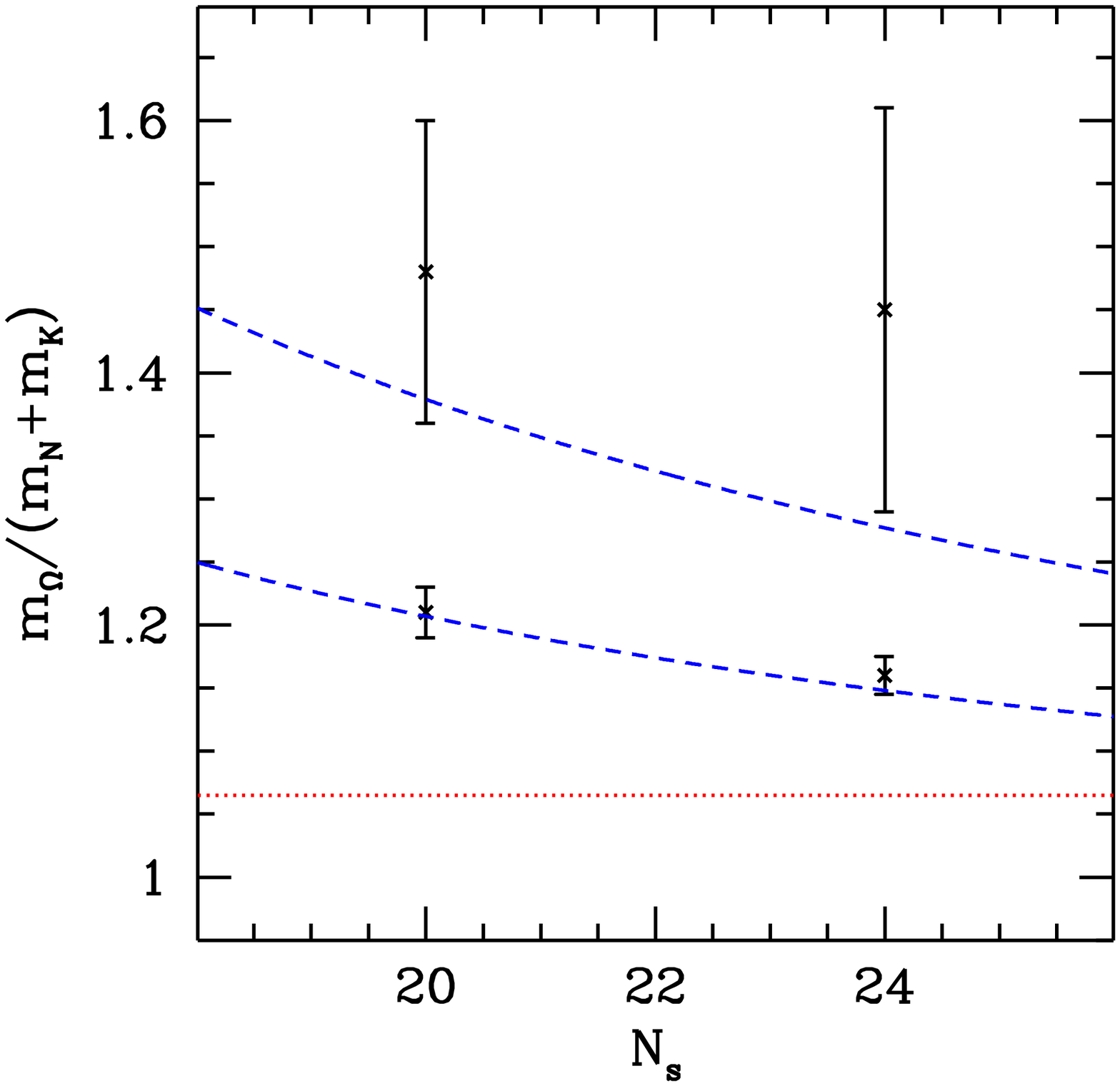}
\end{center}
\caption{Spatial size dependence of energies in the pentaquark channel in quenched 
QCD\protect\cite{fodor}. }
\label{fig:penta}
\end{figure}

In Fig.~\ref{fig:penta} we reproduce the spatial size 
dependence of the energies in the $1/2-$ and $1/2+$ channel calculated 
with five operator basis in quenched QCD\cite{fodor}.  The dashed lines 
indicate the size dependence expected for scattering states, and the 
horizontal dotted lines show the experimental value quoted for the 
$\Theta^+$ state.  As the figure indicates, the authors conclude that 
there is no evidence for the existence of pentaquark states in their data.  

While conclusions are not unanimous among the studies\cite{penta-since,fodor},  
it is our view that the data published to data 
are more consistent with the absence of 
pentaquark states at present.  Caution should be stated that all data 
so far have been generated in quenched QCD, mostly at a fixed lattice spacing 
and at relatively large quark masses.  Truly realistic tests with dynamical 
light quarks with large volume and small lattice spacing require further 
work in the future. 
  
\section{Conclusions}
\label{sec:conclusion}

Currently lattice QCD simulations are in a state of transition.  
Realistic simulations including dynamical up, down and strange quarks 
are becoming routine, and chirally invariant quark actions are beginning to be 
used increasingly frequently.  This trend will accelerate as computers 
with 10 Tflops-class capability, starting with QCDOC and followed by 
ApeNEXT, PACS-CS and also BlueGene/L at several institutions, becomes 
available.  We hope that single hadron properties will be fully understood, 
inviting us to venture into the World beyond, in the near future.

\section*{Acknowledgments}
This work is supported in part by the Grants-in-Aid of the Ministry of 
Education (No. 15204015).

\end{document}